\documentclass[20pt]{article} 

\usepackage{enumerate}
\usepackage{url}
\usepackage{textgreek}
\usepackage{moresize}

\usepackage{geometry} 
\usepackage{graphicx} 

\usepackage[parfill]{parskip} 

\usepackage{amsmath}

\usepackage{booktabs} 
\usepackage{array} 
\usepackage{paralist} 
\usepackage{verbatim} 
\usepackage{subfig} 

\usepackage{fancyhdr} 
\usepackage{sectsty}

\usepackage[nottoc,notlof,notlot]{tocbibind} 
\usepackage[titles,subfigure]{tocloft} 

\title{How NOT to Fool the Masses When Giving Performance Results for Quantum Computers }
\author{Catherine McGeoch \\ D-Wave Quantum (retired)}
\date{\today} 

\begin{document}
\maketitle

\large

\paragraph{\bf Abstract:}
In 1991, David Bailey wrote an article describing techniques for overstating the performance of massively parallel computers.  Intended as a lighthearted protest against the practice of inflating benchmark results in order to ``fool the masses''  and boost sales, the paper sparked  development of procedural standards that help benchmarkers avoid methodological errors leading to unjustified and misleading conclusions.    

Now that quantum computers are starting to realize their potential as viable alternatives to classical computers, we can see the mistakes of three decades ago being repeated by a new batch of researchers who are unfamiliar with this history and these standards.  

Inspired by Bailey's model, this paper presents four suggestions for 
newcomers to quantum performance benchmarking, about how not to do it. They are: (1) Don't claim superior performance without mentioning runtimes; (2) Don't report optimized results without mentioning the tuning time needed to optimize those results; (3) Don't claim faster runtimes for (or in comparison to) solvers running on imaginary platforms; and (4) No cherry-picking (without justification and qualification).  Suggestions for improving current practice appear in the last section.  


\section*{Introduction}
In 1991, David Bailey published a paper titled {\em Twelve Ways to Fool the Masses When Giving Performance Results on Parallel Computers}, 
which illustrated how researchers 
could devise benchmark tests that show their favorite architecture in the best possible light, even though the reported performance would 
never be experienced by anyone outside 
the lab [1, 2].  Offering  tongue-in-cheek suggestions for strategic wrongdoing, the article had a profound impact on the field, prompting 
what was called a ``mini scandal''  at the time [3],  sparking the development of procedural standards for benchmarking high-performance 
computer platforms [4--9], and instilling lasting distrust of computer manufacturers who report on performance of their own products [10].

This phenomenon is not confined to benchmarking computer platforms.  
A large literature has also developed around guidelines for empirical evaluation of algorithms, 
which aims to ensure ``integrity in and reproducibility of the reported results''  [11] (see also [12--23]). 
But sound methodology is not always enough to mitigate distrust of the profit 
motive: in 2018, a respected benchmarking program for exact optimization solvers\footnote{A {\em solver} is an algorithm or heuristic that has been instantiated in code or hardware and can be measured empirically.  An {\em exact} solver guarantees to find optimal solutions to a given optimization problem if given enough time, whereas a {\em heuristic} solver aims to find good solutions quickly in typical cases, but with no guarantees.} saw withdrawal of the leading commercial participants after one company was accused of cherry-picking results for a marketing brochure [24, 25]. 

Now, as quantum computers are starting to realize their potential as viable alternatives to classical computation, the errors that Bailey called out are being repeated by a new generation of benchmarkers who are unfamiliar with these standards and expectations.  Inspired by Bailey's model, this paper presents four suggestions for reporting 
quantum performance results in the technical literature.   They are:
\begin{enumerate}
\item Don't claim superior performance without mentioning runtimes.
\item Don't report optimized results without mentioning the tuning time needed to optimize those results.
\item Don't claim faster runtimes for (or in comparison to) solvers running on imaginary platforms.
\item No cherry-picking (without justification and qualification). 
\end{enumerate}
 
Cautionary examples are taken from the empirical literature on both gate model (GM) and quantum annealing (QA) performance.   
  
I want to emphasize my strong belief that the problematic practices described here were not intended to fool anyone.  For one thing, most were developed by academic scientists with no obvious reason to promote or impede any particular technology.  For another, there is no attempt to 
conceal testing procedures; most authors provide sufficient details that infractions 
are easy to spot.  Instead, difficulties arise because ``performance'' means different things to 
different groups of quantum stakeholders, including physicists, computer scientists, operations researchers, and those engaged in benchmarking commercial platforms and software.  

For example, with commercial benchmarks such as SPEC [26] and LINPACK [27],  the prime 
directive is to measure performance in a way that could be replicated 
by users.  Benchmarking reports are expected to honor this covenant with the 
masses (also known as prospective users):  {\em These results are relevant to, 
and reflective of,  your everyday use of this product.}  

For many physicists, however, the purpose of a quantum performance study is to understand
the nature and capability of quantum computing, which requires building accurate empirical models of the impact of noise and control errors on 
output quality. Research papers about quantum performance are not normally aimed at the masses; instead they are expected to be read by other researchers working to improve quantum technologies.  

Problems arise when a study reports a successful performance outcome for a commercial quantum solver, which is (mis)understood as a contribution to the commercial benchmarking literature.  When scrutiny reveals that the rules were not followed and the covenant was broken, the cynical masses dismiss the paper as yet another failed attempt to hype the product.  
 
In this context, the discussion herein should {\em not} be interpreted as critiquing experimental methodology {\em per se}, but rather as illustrating how best practice in one field can be seen as biased and misleading practice in another.  To avoid awkwardness,  the next sections follow Bailey's precedent by presenting problematic examples without citation.  Yes, the originals are easy to track down nowadays, but the important point is that it doesn't matter who wrote the papers: these authors were following {\em common practice} in their own discipline.  Awkwardness only arises when the work is misunderstood---rightly or wrongly---as intended for a different audience.  The last section contains suggestions 
for mitigating this type of miscommunication going forward.  

\section*{Four Principles} 

\subsection*{Principle 1. Don't claim superior performance without mentioning runtimes.}  

\begin{quotation}
\noindent {\em We commit the {sin of inappropriateness} when we derive a claim that is predicated on some fact that is absent from the evaluation.} --- Stephen M. Blackburn et al. [15] 
\end{quotation}

Classical algorithmic performance analysis is based on this foundational question: {\em How many compute resources are needed to get a correct answer?}  The resource of first importance is {\em computation time}, measured by counting  dominant operations in studies of abstract performance, and by counting microseconds in benchmarking work. 

Optimization problems have the special property that every solution comes with a numerical quality score.  Also, optimization heuristics are often defined in terms of a user-set work parameter $W$ that controls a tradeoff between solution quality and computation time:  we expect increasing work to produce better-quality solutions but also longer computation times.  For these reasons, heuristic performance analysis requires a two-dimensional metric, asking: {\em How much solution quality $S$ does the solver deliver per unit of computation time $T$?} 

In quantum computing, an important research priority is to locate, understand, and limit the effects of hardware errors on solution quality.  In current GM-based research, performance analysis is almost exclusively focused on modeling errors, asking: {\em How closely do outputs from the real-world quantum processor resemble those from an ideal noise-free computation?} 

Proctor et al. [28] present a thoughtful rationale for this mindset, arguing that ``the importance of other limiting factors, such as speed or power consumption, pales in comparison.''  This view is widely shared 
in the research community; for example, popular performance metrics such as 
quantum volume measure output quality but not computation time. 
(The empirical literature on QA performance is rather more balanced, with papers addressing all three of the above questions.) 

This mismatch of questions and expectations creates problems 
when a physics-style research paper evaluates performance of a quantum product (hardware, 
software, or both) in comparison to other solvers rather than to a theoretical ideal.  Consider these two examples: 
\begin{itemize}
\item A recent paper (here referred to as [R1]) compares a heuristic optimization solver running on a GM 
platform (here called GM$_a$), to a simple classical solver (C$_a$), and a D\nobreakdash-Wave annealing platform (QA).  
The abstract states that GM$_a$ outperforms both C$_a$ and QA;  these claims are based on measuring {\em success probability} $\pi$, 
which is the probability finding an optimal solution in an output sample.  Computation times are nowhere mentioned in the 
main text (although they are briefly mentioned in an appendix; see Principle 2).   

\item Another paper [R2]  describes performance of a different quantum optimization 
heuristic running on a GM platform (GM$_b$). The abstract states that GM$_b$
outperforms four alternative solvers (again in terms of success probability): QA, another GM heuristic (GM$_c$), and two 
classical solvers C$_b$ and C$_c$.  Computation times are not mentioned, except to note that QA annealing time was set to $2000$ \textmu s.  
\end{itemize}

From the perspective of classical benchmarking, the claims made in the abstract  are not supported by the experiments and outcomes described in the text.  This is an example of  {\em semantic discord}, a term used by philosophers of science to describe a situation where an apparent disagreement about a given subject is really a disagreement about the meanings of words---in this case ``outperforms''---being used to describe the subject.  

The problem is that {\em all} of the solvers mentioned above are instantiations of heuristics, and each solver $X$ must have been run using a specific work parameter setting $W_X$ to produce observations of solution/time pairs $(S_X,T_X)$.  
These two-dimensional data points create a Pareto frontier\footnote{A Pareto frontier is used to analyze outcomes in multi-objective  test 
scenarios where improving one objective might require sacrificing the other.  Points on the frontier, called Pareto-efficient, represent multiple ways to optimize the trade-off.} of outcomes.  

That is (assuming lower is better in both dimensions),  if {\em both}  $S_X< S_Y$ and $T_X< T_Y$, then we can conclude that  
$X$ outperforms $Y$.  But if $X$ only dominates in one dimension (e.g.,  $S_X < S_Y$ but $T_X > T_Y$),  then both points are on the frontier, 
and neither dominates the other.  No further conclusions can be drawn about comparative performance, since it is possible that $X$ 
found better-quality solutions {\em only because} it was granted extra computation time.

Without information about $T$, it is impossible for the reader to assess whether the 
experimental results indeed support claims of superior performance, or are simply artifacts of the test design.  Readers cannot attempt to replicate 
these results because critical information about $W$ and $T$ is missing.

This issue applies not just to optimization problems, but to any study of quantum 
performance  that involves a tradeoff between computation time and solution sample quality. If solver $X$ has high 
probability $\pi_X = 0.6$ of finding the correct solution, and runs in $T=1$ second, we can expect to find it within 2 seconds.  If solver $Y$ 
has low success probability $\pi_Y = 0.1$, and runs in $T=1$ millisecond, we expect to find it in 10 milliseconds. 
Without information about runtimes, how do we know which one is more efficient?  

\subsection*{Principle 2. Don't report optimized results without mentioning 
the tuning time needed to optimize those results.} 
\begin{quotation}
\noindent{\em While performance measurements might seem objective on the surface, there 
are many different ways to influence benchmark results to favor one system over the other, 
either by accident or on purpose.}  --- Mark Raasveldt et al. [9] 
\vspace{.1in}

\noindent{\em It goes without saying that comparisons of tuned versus untuned algorithms are not fair and should be avoided.}   ---  Thomas Bartz-Beielstein and Mike Preuss  [13] 
\vspace{.1in}

\noindent {\em If different parameter settings are to be used for different instances, the adjustment process must be well-defined and algorithmic, the adjustment algorithm must be described in the paper, and the time for the adjustment must be included in all reported running times.}  --- David  S. Johnson [19]
\end{quotation}

The question of  fair testing  is much discussed in the classical  benchmarking literature.  The general idea is to uphold the covenant 
with the masses by not reporting  performance results that require heroic 
tuning efforts.  At the same time, the manufacturer has a reasonable desire to show off the product in its best light.  
That being said, some users are interested in best-case performance of solvers that can be specialized to particular types of inputs; 
others are interested in robust---best worst-case---performance of solvers that do not require (much) tuning effort.  

Fair-test procedures have been developed for computer performance 
benchmarking to balance these conflicting desires. For example, benchmarks of commercial 
hardware often report both {\em peak} (optimized) and {\em base} (default) performance, 
presenting a range of possible outcomes rather than a single number.  

In optimization, fair-test guidelines have been developed for {\em instantiating} 
(assigning values to) the several user parameters (such as $W$) that typically come 
bundled with heuristic solvers.  This step is crucial, because the no free lunch (NFL) principle\footnote{The NFL {\em theorem} (not discussed here) applies to a specific class of optimization algorithms;  the NFL {\em principle} applies informally to classical heuristics  with multiple parameters, which use instantiated strategies to search the solution space created by a given input.}   in optimization [29, 30] tells us that the outcome of a benchmark test is governed primarily by the agreement between input structure and solver structure, as determined by its instantiated parameters.  To avoid a type 
of ``overfitting'' of parameters to inputs, fair-test guidelines require that all solvers use default parameter settings throughout, or receive equal tuning effort in a preprocessing step before fixed-parameter tests begin;  otherwise, if instance-specific tuning is used, it should be algorithmic and included in total computation time [9, 12, 13, 19, 29, 30, 31].  

The NFL principle tells us that an observation that solver X outperforms solver Y on input set A cannot be safely generalized to 
different instantiations of X and Y on the same inputs. Therefore, benchmarking reports should describe tuning policies with enough detail 
to support replication of the results,  and claims about performance that extend beyond the test boundaries should be avoided [30--32].    

Fair testing is not always emphasized in physics-style performance analysis, where the more 
common research objective may be to demonstrate that a phenomenon exists, not that it is common nor easy to find.  
Extreme tuning efforts are routinely applied and tuning time is rarely mentioned.   

For example, the QAOA\footnote{The Quantum Approximate Optimization Algorithm, also known as the 
Quantum Alternating Operator Ansatz, is the most-studied optimization heuristic for gate model platforms.}  heuristic requires instantiation of (at least) three parameters, the iteration count $p$ and two vectors $(\gamma_i, \beta_i)$ with settings for each iteration $i = 1 \ldots p$.  The problem of 
optimizing $(\gamma, \beta)$ has been proven NP-hard [33], and exhaustive parameter sweeps   
to find good values of $(\gamma,\beta)$ are sometimes necessary for QAOA to be effective.  Most QAOA performance 
studies report solution quality based on the final, optimized circuit, omitting to mention time needed to find optimal values 
for $(\gamma,\beta)$;  it is not unusual for the tuning time to dominate the final QAOA circuit time by many orders of magnitude.  
 
This common practice doesn't normally attract attention outside the field. 
But consider how the two papers from Principle 1 can be interpreted: 
\begin{itemize}
\item The paper [R1] devotes much of the text to describing the 
parameter optimization procedure used for GM$_a$.  An appendix lists the parameter settings 
selected for the other solvers, with no explanation of how or why they were chosen.  

The appendix also contains a more detailed but incomplete discussion of computation 
times for GM$_a$ and QA.  For example, on one set of inputs, the GM$_a$ solver performed 16 optimization steps per instance, resulting in a pool of 1.4 million solution samples that took about 20 minutes of wall-clock time to be generated.  However, the {success probabilities} $\pi$ used in the main paper are based on much smaller pool sizes:  for one instance that is discussed in detail, the pool size is 3288, which presumably reflects the last, fully optimized, run of the circuit.      

The paper borrows QA performance results from a third-party publication [34], which states 
that each input required on average 1.3 million samples and 22 minutes of wall-clock time.  
However, the authors apparently misread this statement as describing tuning time, analogous to that needed to optimize $(\beta, \gamma)$.  
In fact, it refers to the total time needed to 
generate the dataset for analysis, not the time actually needed to solve an input.  
A response paper by D-Wave authors explains that in fact QA needs on average 500 samples and about 0.5 seconds of wall-clock time per input, using default (unoptimized) parameters. Recalculating $\pi$ with correct sample sizes (500 versus 1.3 million) boosts QA success probabilities enough 
to reverse the claim of higher success probabilities from GM$_a$. 

\item The paper [R2] contains a list of parameter settings used for each solver, and states that success probabilities for GM$_b$ are based on the final iteration of a 10-step parameter optimization process.  There is otherwise no mention of what the optimal parameter settings were, nor how much time was needed to find them. 
\end{itemize}

From the perspective of fair-testing standards in classical benchmarking, both papers fall short of expectations.  There is no discussion 
of tuning policies, nor (apparently) any attempt to apply 
equal tuning efforts.  Runtimes, when mentioned, do not match the data.  Readers are not able to assess whether reported performance is merely an artifact of the choice of test parameters, nor how much effort would be required to replicate the results, nor which solver would prevail were tuning times properly accounted for. The cynical masses, trained to expect fair-test shenanigans in commercial benchmarking, find their cynicism justified.  

\subsection*{Principle 3. Don't claim faster runtimes for (or in comparison to) solvers running on imaginary platforms.} 

\begin{quotation}
\noindent {\em The practice of linearly extrapolating one's performance results to a larger system is doubly perplexing because the question of whether various computer designs and applications will ``scale''
is in fact an important topic of current research in the field of parallel computing.}  --- David H. Bailey [2]
\vspace{.1in}

\noindent {\em Studies that try to extrapolate asymptotic running time by studying times of instances of [small size] can often be led astray.}  --- David S. Johnson [19] 
\end{quotation} 

Studying asymptotic performance on an abstract model of computation is the bread-and-butter of 
theoretical algorithm analysis. The primary time metric, also abstract, is based on identifying 
a core (i.e., asymptotically-dominant) operation and counting the number of times 
it is performed by a given algorithm or heuristic, with the goal of finding a function that 
relates core operation counts to problem size $N$.  

Sometimes in QA performance studies, stopwatch-style runtimes are assigned to core operations.  For example, a collection 
of research papers studying a property called {\em quantum speedup}
(e.g., [35]) use the time-to-solution (TTS) metric, which combines a measurement of time $T$ per core operation with a statistic $\overline{C}$ for the number of core operations needed to find an optimal solution in an iterative search of the solution space.  Quantum speedup studies tend to 
underreport classical computation times, because TTS ignores computation costs outside of the 
core operation, and in some cases because TTS is arithmetically adjusted (e.g., divided by $N$) to study performance scaling on an 
abstract parallel model of computation.  

Research papers on quantum speedup are not normally mistaken for benchmark-style 
performance studies because the analysis focuses on {\em shapes} of TTS curves,  rather than on absolute computation times. 
However, this approach is sometimes taken up by researchers who adapt TTS to other purposes. 

One problematic practice is extrapolating TTS curves measured on small existing platforms to predict runtimes on large future platforms.  Experts in 
empirical study of classical algorithms warn that extrapolating computation {times} without reliable models is 
never a good idea: see [2, 19, 36, 37] for examples of predictions gone awry.  Extrapolation in quantum benchmarking is arguably more hazardous, for two reasons: 

\begin{itemize}
\item {\em Parameterized solvers are not suited for extrapolation.}  Paper [R3] compares a classical device (C$_d$) that performs annealing-style computations to a D-Wave processor (QA).  The abstract states: ``On instances with over 50 vertices, a several orders of magnitude time-to-solution (TTS) difference exists between [C$_d$] and [QA].  An optimal annealing time analysis is also consistent with a significant projected performance difference.''  

The text contains a proper caution to readers against naive extrapolation to larger problem sizes.   This caution is followed by discussion 
of a table containing measured data up to $N=(60, 55, 100)$ and extrapolated data up to $N=(100, 100, 200)$, for three input sets.  The claim that 
C$_d$ outperforms QA holds on the first two input sets (not the third), and indeed C$_d$ outperforms QA by very large margins 
on the extrapolated data: the largest QA computation time mentioned is $10^{19}$ seconds, which would surely require an imaginary quantum
platform.\footnote{For comparison, about $10^{17}$ seconds have elapsed since the Big Bang.} 

Even if the imaginary experiment could be performed, a solver running at large $N$ using a work parameter $W$ that has been
optimized for small $N$ is unlikely to exhibit well-tuned performance.  That is, the slope of the {measured} TTS curve is approaching an 
unknown asymptote at an unknown rate that depends critically on the work parameter $W$.  There is no known model of how $W$ must scale with $N$ to ensure good performance;  if it were known, experiments likely wouldn't be needed. 

\item {\em Effects of technological improvements on performance of future quantum hardware cannot be predicted.}   A paper [R4] studying a QA effect known as {finite-range tunneling} reports on performance comparisons to two classical algorithms (C$_e$ and C$_f$) that do not take advantage of tunneling. They remark that, based on previous experience with a D-Wave Two\texttrademark{} platform (500+ qubits), they predicted that the D-Wave 2X\texttrademark{} platform (1000+ qubits) would be about $10^4$ times faster than C$_e$; in fact it was more than $10^8$ times faster, due to improved noise suppression and a colder fridge. They write:  
\begin{quotation} 
\noindent {\em For this reason, the current study focuses on runtime ratios that were actually measured on the largest instances solvable using the current device, rather than on extrapolations of asymptotic behavior which may not be relevant once we have devices which can attempt larger problems.}
\end{quotation} 
This property of performance improvement by generation has been demonstrated across all five hardware upgrades announced by D-Wave, 
and is expected to continue.  
\end{itemize} 

Listen to the experts: extrapolation-based predictions about future performance of heuristic solvers running on classical or quantum platforms can turn out to be very wrong.  Skeptical readers, familiar with benchmarking guidelines designed to avoid this pitfall, tend 
to dismiss extrapolated results as either naive or deliberately inflated to impress the masses.    

\subsection*{Principle 4. No cherry-picking (without justification and qualification).} 

\begin{quotation} 
\noindent {\em Pet Peeve 11:  The one-run study. Unless a study covers a wide enough range of 
instances..., the conclusions drawn may well be {\em wrong}, and hence irreproducible for that reason.... It can be dangerous to infer too much from a single run on a single instance.}  --- David S. Johnson [19] 

\vspace{.1in} 
\noindent {\em It is desirable for the suite to contain a wide variety of problems with different characteristics.  In this way, a good problem suite can be used to highlight the strengths and weaknesses of different algorithms}. --- Thomas Bartz-Beielstein et al. [14]

\vspace{.1in} 
\noindent 
{\em Nearly every problem set inspires complaints about its bias and limited scope.... It is unclear that we would even be able to recognize a representative problem set if we had one.} --- John N. Hooker [18] 

\end{quotation} 

As mentioned previously, the NFL principle tells us that solver X outperforming solver Y on input set A does not imply that X would outperform Y under different instantiations.  Furthermore, it does not imply that under the same instantiations, X would outperform Y on different input sets B, C, or D.  In the big picture, generalizing benchmarking outcomes to {\em anywhere} outside the scope of testing is fraught with danger.  

One relatively safe way to improve generality is to expand the scope of inputs (and comparison solvers) used in the benchmark study.  Thus, standard benchmark repositories for optimization solvers typically contain thousands of instances from broad varieties of problem categories [38--42].  Ideally, such a repository can be used to {\em characterize} performance by developing empirical models that predict which solvers are best suited for which categories of problems.    

Unfortunately for quantum benchmarkers, nearly all classical repositories contain only inputs that are much too large to fit on current quantum platforms. Standardization of quantum-specific repositories would be premature at present, given our current lack of knowledge about what tasks quantum computers are (or will be) good at.  For now, quantum researchers must be left to their own devices when selecting inputs and comparison solvers for study.  

Of course, small test scope is not incompatible with good research: single-instance tests can be
valuable for close analysis of statistical properties, or for proof-of-concept illustrations of new ideas.  Simple parameterized inputs (which don't necessarily resemble real-world inputs) can be useful for building models of newly-discovered algorithmic mechanisms.  Comparison solvers may be selected simply because they are handy for illustrating certain technical points, not because they represent the best-available competition for racing purposes.  

Research time and page counts being finite, every benchmark test design requires judicious selection of what to include and what to 
leave out.  Hooker [18] argues that large community-built problem repositories tend to co-evolve with successful solvers, eventually becoming biased against new approaches that may have strengths complementary to the accepted canon.  By this argument,  cherry-picking---inadvertantly selecting inputs that uniformly favor one solver over another---is an unavoidable hazard of competitive benchmarks, even when they are designed with the best intentions.  

Although cherry-picking can't be avoided (and may be impossible to detect), the masses rightly hold benchmarking reports to high standards when 
conclusions are drawn about the implications and significance of reported results, asking: {\em Does the generality of performance claims in the paper match the scope of the experiments?}  

Hence the parenthetical qualification in the statement of this principle.  Answering this question 
requires clear justification of how and why test components (inputs and solvers) were selected to meet 
the goals of the study, and explicit discussion of limits on what can be concluded from the outcomes.   

Suspicion peaks when these explanations are missing and the authors of what appears to be a benchmarking-style report are affiliated with the company that produced the winning solver.  If the paper contains broad claims based on miniscule or contrived input sets or comparisons to straw-man solvers, the 
masses are more likely to suspect dark commercial interests than imprecise or roseate wording as the root cause.  
Consider these examples: 

\begin{itemize} 
\item In paper [R1], the abstract states that GM$_a$ outperforms both C$_a$ and QA, 
and furthermore that these results, ``demonstrate the first time a gate-model quantum computer has been  able to outperform an annealer.''  The claimed superior performance of GM$_a$ over QA is based on exactly {one} input instance, out of three used for comparison (the other two yielded ties).  QA performance numbers were taken from a third-party paper that describes tests on a pool of 20 instances; the remaining 17 instances are not mentioned in [R1]. All of the authors are from the company that built solver GM$_a$.

\item In paper [R2], performance claims for GM$_b$ (compared to QA, GM$_c$, C$_b$ and C$_c$) are based on exactly two input instances generated by the authors. The abstract states that the result ``can be considered as the start of the commercial quantum advantage era.''  All five authors are affiliated with the company that developed the quantum software (two also have university affiliations). 

\item Quantum speedup papers typically compare quantum and classical solvers using 
exhaustive ``sweeps'' of possible parameter combinations (always $W$ and sometimes others).  
Computation times on scales of days may be needed to generate a pool of outcomes for each solver, from which best results are selected {post hoc}.  These results forms a lower-bound envelope of measured TTS curves, with runtimes corresponding to a single optimized trial. This practice 
is justified by the research goal of measuring performance of all solvers under best-tuned conditions.  From the perspective of classical benchmarking, 
however, this a type of data cherry-picking, since users could not expect to observe best-quality solutions without first building the large pool to draw from.  

Paper [R3], adapting this approach, describes three post-hoc techniques for filtering bad outcomes for solver C$_d$, in their comparison of  TTS curves for C$_d$ and QA.  At the time of publication, solver C$_d$ was a recently-announced commercial product;  the author list contains both academic and industry researchers, the latter from the company that developed C$_d$.  

\item  Paper [R4] compares a QA solver against classical solvers C$_e$ and C$_f$ (not commercial products), using inputs from a single problem 
class invented by the authors, and reporting very large speedups of QA over both.  At the time of publication, industry bloggers accused them of 
cherry-picking the competition by not including the best-available classical solver (C$_g$) in their tests.  In the paper, the authors acknowledge that C$_g$ would likely perform better than QA (it did), but argue that it is too narrowly specialized to specific types of inputs to qualify as a viable general-purpose competitor (which turned out to be the case;  C$_g$ is no longer studied).  All of the authors work in the quantum industry, but not at the same company that manufactures QA  platforms.  
\end{itemize}  
It is always possible that the outcomes of any performance study might 
be due to inadvertent cherry-picking.  What about the above cases:  were results deliberately cherry-picked 
for deceptive purposes?  Does knowing the authors' affiliations affect your decision?  

No matter where your answers to these questions might land, I believe that suspicions could have
been greatly allayed if the authors had paid more attention to justifying their test designs and properly qualifying the scope of their conclusions.  Some suggestions for improving current practice in publication of quantum performance papers appear in the next section.

\section*{Some Modest Suggestions} 
\begin{quotation}
\noindent{\em  Manufacturers commonly report only those benchmarks (or aspects of benchmarks)  that show their products in the best light.  They also have been known to mis-represent the significance of benchmarks, again to show their products in the best possible light.  Taken together, these practices are called {\em bench-marketing}.} --- Wikipedia [10] \vspace{.1in}

\noindent {\em Those who cannot remember the past are condemned to repeat it.} \\--- George Santayana [43] 
\vspace{.1in}

\noindent {\em What we've got here, is failure to communicate.} --- Captain, in {\em Cool Hand Luke} [44]  
\end{quotation} 
 
In the big picture, it doesn't matter who is to blame for possible misunderstandings of authors' intentions. 
The concern, of course, is that a critical  mass of {\em perceived} hype could spark widespread distrust of the quantum computing industry 
and trigger a quantum winter, or perhaps something like downscaling of the parallel computing industry in the years following publication 
of Bailey's Twelve Rules [45, 46].  (Please remember, however, that correlation is not causation.)  

The more important question is how to repair the current situation and avert those potential outcomes. 
To those ends, I offer some suggestions for improving communication between scientists and the nonexpert masses, 
also known as potential customers, investors, funding agents, industry journalists, and other quantum stakeholders. 

 {\bf For quantum researchers.}  Be aware that there is a large external community that craves information about progress in quantum computing and may be looking for answers in your paper.  
Authors who are affiliated with commercial organizations must be prepared for extra scrutiny of any publication that contains performance comparisons. 
\begin{itemize}
\item If your paper {\em is not} intended to be read as a commercial benchmarking study, choose your words carefully in the abstract, introduction, and conclusions sections, to avoid giving that impression.  It helps to clearly define exactly what you mean by performance, and to explicitly 
qualify the implications of your results.  

\item If your results {\em are} intended to reach the masses, consider doing some background reading on classical 
benchmarking protocols and expectations; papers marked with {\bf *} in the reference list are a good place to start.  The short version:  A claim that X 
outperforms Y needs to be supported with a clear definition of performance, disclosure of test designs and tuning policies sufficient to support 
replication,  justification of choices made in defining the project scope, and qualification regarding limits of your conclusions.  
\end{itemize}

{\bf For journal editors and peer reviewers.}  Presumably the paper was submitted to the journal because it is intended to advance scientific knowledge (something not normally expected of benchmarking reports, which aim to report observations without necessarily explaining them). 
In addition to helping authors meet the usual high standards for scientific research publication, please help them
frame their discussion of results and research contributions to avoid misunderstandings. 

{\bf For the masses.} Be aware that there is a  large community of researchers, focused on modeling and 
understanding performance of quantum computers, who are not thinking about the questions you have.  Words like ``performance,''   ``benchmarking,''  and  even ``computation time'' in these papers may have very different meanings from what you expect.  Your patience is requested as the field works toward broader awareness of these issues, better consensus on terminology, and greater attention to clarity of goals and thoughtful interpretation of results. 

\subsection*{Acknowledgement}
I thank David H. Bailey for several helpful suggestions. 

\section*{References}
{\em Papers marked with {\bf *} form a beginner's reading list on benchmarking guidelines.} 


\end{document}